\documentclass[twocolumn]{aastex631}

\usepackage{booktabs}
\usepackage{array}
\usepackage{graphicx}
\usepackage{natbib} 
\usepackage{amsmath,amssymb} 
\usepackage{color} 
\usepackage{xcolor} 
\usepackage{hyperref} 
\hypersetup{linkcolor=cyan,citecolor=magenta,filecolor=yellow,urlcolor=blue} 

\shorttitle{Islands of Electromagnetic Tranquility}
\graphicspath{{./}{image/}}

\begin{document}

\title{Islands of Electromagnetic Tranquility in Our Galactic Core and Little Red Dots that Shelter Molecules and Prebiotic Chemistry}

\email{yu.wang@icranet.org, ruffini@icra.it}

\author[0000-0003-0829-8318]{R.~Ruffini}
\affiliation{ICRANet, Piazza della Repubblica 10, 65122 Pescara, Italy}
\affiliation{ICRA, Dip. di Fisica, Sapienza Universit\`a  di Roma, Piazzale Aldo Moro 5, I-00185 Roma, Italy}
\affiliation{ICRANet-AI, Brickell Avenue 701, Miami, FL 33131, USA (in preparation)}
\affiliation{Universit\'e de Nice Sophia-Antipolis, Grand Ch\^ateau Parc Valrose, Nice, CEDEX 2, France}
\affiliation{INAF,Viale del Parco Mellini 84, 00136 Rome, Italy}

\author[0000-0001-7959-3387]{Yu Wang}
\affiliation{ICRA, Dip. di Fisica, Sapienza Universit\`a  di Roma, Piazzale Aldo Moro 5, I-00185 Roma, Italy}
\affiliation{ICRANet, Piazza della Repubblica 10, 65122 Pescara, Italy}
\affiliation{ICRANet-AI, Brickell Avenue 701, Miami, FL 33131, USA}
\affiliation{INAF -- Osservatorio Astronomico d'Abruzzo, Via M. Maggini snc, I-64100, Teramo, Italy}

\begin{abstract}
Both the Galactic Center and little red dots (LRDs) host million-solar-mass black holes within dense, cold reservoirs of molecules associated with dust grains, and are electromagnetically tranquil. These conditions enable complex molecular chemistry and may serve as natural laboratories for prebiotic genetic evolution by allowing the synthesis of organic molecules essential for life.
\end{abstract}

\section{Introduction}

Over the past six decades, our understanding of Galactic Centers has evolved from the discovery of luminous compact sources to a deeper perspective shaped by the Milky Way's quiet nucleus. Sagittarius A$^*$ exemplifies a central compact core of about $10^6\,M_\odot$ accreting at a very low rate, surrounded by the Central Molecular Zone (CMZ), a region rich in cold molecular gas \citep{2022ApJ...930L..12E}. More recently, JWST observations have identified ultra-compact, dusty ``little red dots (LRDs)'' at high redshift \citep{2024ApJ...963..129M}, which also appear to host black holes of similar mass embedded within dense, molecule-rich cores. Like the Milky Way's center, these proto-galaxies are bright in the infrared but faint in X-rays, suggesting a comparable environment. This raises the central question of our article: since the CMZ of the Milky Way hosts molecular clouds containing complex prebiotic molecules, could similarly tranquil regions around million-$M_\odot$ black holes in LRDs also support the formation of complex molecules and prebiotic chemistry, potentially making them ``Islands of Electromagnetic Tranquility'' both in our Galaxy and in the LRDs of the distant universe?

\section*{Low Radiation Milky Way's Core}

The center of our Milky Way hosts a compact object of $4 \times 10^6 M_\odot$, Sagittarius A* (Sgr~A*), which is extremely under-luminous and quiescent compared to the active galactic nuclei (AGN) seen in many other galaxies. Sgr~A* is accreting very little material at present, radiating at only a tiny fraction (below $10^{-9}$) of its Eddington luminosity. Unlike a quasar or a Seyfert nucleus, the Milky Way's core emits no powerful jets or intense X-ray outbursts, because it lacks the existence of a fast rotating Kerr black hole and the large, sustained infall of gas required to ignite the AGN phase \citep{2022ApJ...930L..12E, ruffini2025role}. 

From an observational standpoint, our Galactic Center contrasts sharply with a classic AGN. Instead of a brilliant point-like nucleus dominating the galaxy's light, the Milky Way's center is observable mainly via its dense interstellar clouds and star clusters. It is rich in molecular gas and dust but comparatively dim in high-energy output. Astronomers have indeed noted that the central region of the Milky Way, despite containing Sgr~A* and a dense stellar cluster, lacks the strong ionization and excitation features that an active nucleus would produce. For example, in external galaxies with AGN, one often sees high-excitation emission lines and ionized cones; by contrast, our center's emission-line regions are mostly low-excitation and associated with star formation (e.g. rings of H\,\textsc{ii} regions) \citep{2017NatSR...716626B}. Kinematically, the Milky Way's bulge and inner disk rotate in an orderly way, with no signs of the violent disturbances that a recent major accretion event might imprint. All these observational clues reinforce that the Milky Way is an inactive galaxy in terms of nuclear activity. In fact, evidence suggests Sgr~A* has been quiescent for at least the last few million years, though tantalizing clues like the giant gamma-ray Fermi bubbles indicate it flickered on in the past (likely a short AGN episode a few Myr ago) \citep{2012MNRAS.424..666Z}. Overall, our Galactic core today is a relatively tranquil environment in the electromagnetic sense, certainly compared to a typical quasar or radio galaxy.

From a theoretical perspective, dark matter likely plays a key role in maintaining the low-temperature, stable core of the Milky Way. The galaxy resides in a massive dark matter halo whose properties, especially the initial angular momentum (spin), influence central gas concentration \citep{2019MNRAS.488.4801J}. Recent JWST studies suggest that low-spin halos allow gas to collapse centrally, forming compact galaxies \citep{2025arXiv250603244P}, while higher-spin halos favor extended disks. The Milky Way’s halo probably had moderate spin, resulting in a large disk and spiral structure, but still provides a deep gravitational potential that gradually channels gas inward. Gas flows along the bar and dust lanes, accumulating in the CMZ ring around Sgr~A* \citep{2019MNRAS.484.1213S}. Because dark matter dominates the central potential in addition to the black hole, much of the gas can remain stable in the CMZ, orbiting at $\sim100$~parsecs, without being rapidly accreted. Thus, under the dark matter framework, a quiescent Galactic center with substantial cold molecular gas and a starved SMBH is expected, as observed in the Milky Way.

\begin{deluxetable*}{lccccccccc}
\tablecaption{Comparison of physical conditions relevant for molecular and prebiotic chemistry. Dust and gas characteristics refer to the cold component on $\sim$100 pc scales.\label{tab:comparison}}
\tabletypesize{\footnotesize}
\setlength{\tabcolsep}{3.5pt}
\tablewidth{0pt}

\tablehead{
\colhead{Objects} &
\colhead{Radius} &
\colhead{$M_{\rm BH}$} &
\colhead{$M_{\rm star}$} &
\colhead{$M_{\rm mol}$} &
\colhead{$M_{\rm dust}$} &
\colhead{$T_{\rm dust}$} &
\colhead{$T_{\rm gas}$} &
\colhead{Ultraviolet} &
\colhead{X-ray} \\[-2ex]
\colhead{} &
\colhead{(pc)} &
\colhead{($M_\odot$)} &
\colhead{($M_\odot$)} &
\colhead{($M_\odot$)} &
\colhead{($M_\odot$)} &
\colhead{(K)} &
\colhead{(K)} &
\colhead{} &
\colhead{}
}

\startdata
Milky Way CMZ & $\sim100$--$250$ & $4\times10^{6}$ & $\sim10^{7}$ & $\sim10^{7}$ & $\sim10^{5}$ & 15--40 & 50--200 & Subdominant & Weak \\
LRD & $\sim30$--$200$ & $10^{6}$--$10^{8}$ & $10^{6}$--$10^{9}$ & $10^{6}$--$10^{8}$\tablenotemark{a} & $\lesssim10^{6}$ & $\lesssim50$\tablenotemark{a} & 50--200\tablenotemark{a} & Subdominant & Undetected \\
\enddata

\tablenotetext{a}{For LRD, $M_{\rm mol}$, $T_{\rm dust}$, and $T_{\rm gas}$ are inferred in this article, there are no direct observations yet.}

\end{deluxetable*}

\section*{Comparing LRDs with Milky Way CMZ}

The analogy between the Milky Way CMZ and LRDs is summarized in Table~\ref{tab:comparison}.

LRDs are tiny, reddish proto-galaxies discovered in the early universe, as early as a few million years after the Big Bang in JWST deep surveys \citep{2024ApJ...963..129M}. They are extraordinarily compact, typically only $30$–$200$ parsecs in radius, which is comparable to our Galaxy’s CMZ size. Despite their small size, many LRDs appear to host massive central black holes of a few million solar masses, similar to SgrA*; surprisingly, the black hole can account for a few tens of percent of the entire stellar mass in these systems. Thus, while both the Milky Way and LRDs have SMBHs in a few million solar masses range, the black hole is a much larger fraction of an LRD than in our Galaxy. Therefore, structurally, an LRD can be viewed as a proto-galaxy that is almost entirely ``core’’ \citep{2024ApJ...963..129M} of the Milky Way, corresponds to just the innermost bulge where the CMZ and nuclear star cluster reside. 

Another major parallel is that both the Milky Way's center and LRDs host active cores without the high-energy output typical of AGN. Sgr~A* currently accretes at a very low rate, forming a radiatively inefficient accretion flow that produces little X-ray or ultraviolet emission; only weak X-ray flares and modest radio/infrared output are observed. Remarkably, LRD galaxies show similar behavior, no X-ray emission has been detected from LRDs to date \citep{2025arXiv250509669S}. It also lacks the observations of radio jets and the extended ionization cones that are produced by AGN-driven winds and collimated radiation \citep{2025A&A...693L...2P}, and their variability of emission is minimal compared to standard AGN \citep{2025A&A...698A.227F}.  This raises questions about the true power source: the central black holes could be heavily obscured or intrinsically accreting at a very low Eddington ratio. In either case, the region around the black hole in LRDs is not dominated by strong high-energy emission, resembling the ``electromagnetically tranquil'' environment of Sgr~A*. LRDs also exhibit gas rotation up to $\sim1000~\mathrm{km\,s^{-1}}$ \citep{2025A&A...697A.189D}, similar to velocities near Sgr~A*, indicating a massive gravity source but without strong winds. In contrast, a quasar core would produce intense X-ray, ultraviolet, and radio jets. Powerful X-ray irradiation establishes extensive X-ray Dominated Regions (XDRs) extending hundreds of parsecs to several kiloparsecs, permeating the nuclear interstellar medium and leading to the efficient dissociation of molecular species \citep{2014ApJ...788...54G}.

Both the Milky Way’s core and LRDs contain large reservoirs of molecular material. In the Galactic Center, the CMZ contains $\sim 10^7 M_{\odot}$  molecular gas and dust \citep{2025A&A...697A..42R}. In LRDs, \textcolor{red}{direct} observations of molecular mass are not yet available, because current data do not spatially resolve these systems nor detect their cold ISM tracers \citep{2025ApJ...990L..61C}. The spectra obtained by JWST mainly constrain the stellar mass and the mass of the central black hole. These observations show that the stellar mass in an LRD is typically within the same order, or one order of magnitude larger than the black hole mass \citep{2025ApJ...983...60C}. This is very different from local massive galaxies, where the stellar mass exceeds the black hole mass by three to four orders of magnitude \citep{2015ApJ...800...20G}. In high redshift star-forming galaxies, in general, the stellar mass and the molecular gas mass are often comparable \citep{2016ApJ...820...83S}. Therefore, if LRDs follow the same mass relation as other high redshift systems, it is reasonable to infer that their total molecular gas mass lies in the range $M_{\rm mol}\sim10^{6}$--$10^{9}\,M_{\odot}$. This range is consistent with the molecular mass contained in the Milky Way CMZ, which hosts in the order of $10^{7}\,M_{\odot}$ of molecular gas. In the CMZ, nearly all of the molecular mass is in the gas phase, and only $\sim10^{5}\,M_{\odot}$ is in dust, corresponding to a gas-to-dust ratio of order $100$, which is typical of a metal-rich environment \citep{2021MNRAS.501.2573D}. In contrast, early universe LRDs likely have substantially lower metallicity, which implies higher gas-to-dust ratios and therefore smaller dust masses at fixed gas mass. If we adopt gas-to-dust ratios $M_{\rm gas}/M_{\rm dust}\sim 300-1200$ for high redshift systems \citep{2025ApJ...993L..40S}, then the dust masses would fall in the range $M_{\rm dust}\sim10^{3}$--$10^{6}\,M_{\odot}$. This indirect estimate is consistent with current observational upper limits. Multiple ALMA programs have carried out deep $1.2$--$1.3\,{\rm mm}$ continuum stacking analyses of LRD samples and found no detections, indicating $M_{\rm dust}\lesssim10^{6}\,M_{\odot}$ \citep{2025ApJ...990L..61C}.

Dust is the main site where molecular complexity can build up, while gas provides the dominant mass reservoir and transport. Prebiotic molecules form and accumulate most efficiently when dust grains are within $\approx 10-50~\mathrm{K}$, cold enough for ice mantles to remain stable, and warm enough for radical diffusion and hydrogenation reactions on the surface \citep{2009ARA&A..47..427H}. The Milky Way CMZ has dust temperatures of $15$--$40$~K and gas temperatures of $50$--$200$~K \citep{2005A&A...429..923R}. Direct observations of cold molecular temperatures in LRDs are very challenging for current telescopes, as these objects are simply too distant and faint, to date there is no confirmed millimeter detection. Therefore, we estimate these properties from theoretical expectations and indirect evidence.

The thermal state of dust and gas at radii of order 100 pc in LRDs can be constrained. First, the observed optical dominated spectrum suggests that the escaping radiation field is redder than in unobscured AGN \citep{2025arXiv250905434G}, raising the possibility that the ultraviolet flux incident on the surrounding medium can be reduced and that the volumetric heating rate of dust is therefore lower. Second, the absence of X-ray emission indicates either intrinsic low luminosity or heavy absorption. If it is intrinsic, as supported by a recent study that utilized a direct dynamical mass measurement of the black hole in an LRD (A2744-QSO1) inferring an accretion luminosity down to 1\% of the Eddington ratio \citep{2025arXiv250821748J}. The dust temperature estimation from a radiation source of luminosity $L$ follows $T_{\rm dust} \propto L^{1/6}$, which implies a factor of $\sim 0.5$ reduction in the LRD dust temperature compared with typical AGN. Alternatively, if the nucleus is strongly obscured, ultraviolet photons can be absorbed at small radii and the outer $\sim 10-100$ pc region may experience a softer radiation field, allowing cold, shielded molecular dust to persist. As an empirical reference point, ALMA observations of the $z=6.9$ luminous quasar host J2348-3054 indicate dust temperatures of $T_{\rm dust}\simeq 73$--$88\,\mathrm{K}$ within radii $r\lesssim216\,\mathrm{pc}$ \citep{2025A&A...695L..18M}. Following the above discussions, a subset of the molecular clouds in LRDs could maintain dust temperatures $<50\,\mathrm{K}$ at $\sim 100$ pc, similar to the cold clouds in the Milky Way CMZ. For the gas at $\sim100$ pc, a lower limit can be obtained by assuming dense molecular gas that is thermally coupled to cold dust, so $T_{\rm gas}\gtrsim 50$~K. An upper limit follows if the gas is only moderately coupled to dust and receives additional heating from turbulence or shocks \citep{2016A&A...586A..50G}, in which case the thermal balance allows $T_{\rm gas}$ to exceed $T_{\rm dust}$ by factors of a few, reaching $\sim 100$~K and plausibly up to $\sim 200$~K \citep{2020A&A...642A.166B}. Together, these considerations motivate $T_{\rm gas} \sim 50-200$~K as a reasonable range, while direct confirmation awaits millimeter molecular line excitation measurements.

We need to notice that for LRDs of low metallicity, the reduced dust-to-gas ratio lowers the available dust surface, and may influence the formation of ice species that seed prebiotic molecules. LRDs have no direct molecular measurements existing, so constraints can only be inferred, for example, from low metallicity Magellanic-Cloud hot cores. In the Large Magellanic Cloud hot core candidate ST11, ALMA sets an upper limit of fractional abundance $X({\rm CH_3OH}) < 8\times10^{-10}$, compared to typical Galactic hot core values of $X({\rm CH_3OH}) \sim 10^{-8}$--$10^{-7}$, implying a depletion of $\gtrsim$1-3~dex \citep{2016ApJ...827...72S}. However, this is not universal. In the LMC region N~113, ALMA finds $X({\rm CH_3OH}) = 2\times10^{-8}$ \citep{2018ApJ...853L..19S}, comparable to the low end of Galactic values. These results indicate that prebiotic chemistry can proceed under low metallicity conditions, but the formation efficiency of prebiotic molecules is reduced, with abundances showing large source-to-source variations \citep{2019ESC.....3.2088S}.

Another noteworthy point is that, from structure formation theory and cosmological turbulence models, the first galaxies are expected to begin forming at redshifts $z \sim 20$ \citep{2011ARA&A..49..373B}. At this epoch, the cosmic microwave background temperature is $T_{\rm CMB} \sim 50\,{\rm K}$, which corresponds to the upper dust temperatures that allow efficient grain surface chemistry relevant for prebiotic molecule formation. In this sense, the onset of galaxy formation occurs at the same cosmic epoch when the background radiation temperature first drops to values compatible with prebiotic molecular assembly. Both of these timescales are linked to the recombination time, which is further related to the values of cosmological parameters.

\section*{Tranquil Cores as Cradles for Molecules and Ingredients of Life}

From the above discussion, the ``electromagnetic tranquility'' LRDs seem to provide favorable environments for prebiotic molecule formation. First, the existence of a large amount of dust and high-density gas supports grain-surface chemistry. Second, a subset of the dust at radii of order tens to $\sim 100$~pc may remain cold ($T_{\rm dust}\lesssim 50$~K), enabling complex surface reactions. Third, intrinsically weak emission or strong internal shielding leads to weak ultraviolet and X-ray radiation, which supplies the energy input needed to drive chemistry without destroying complex molecules. Fourth, such LRD configuration may plausibly persist on timescales of $\gtrsim 10^{5}\text{--}10^{6}$~yr, considering the Salpeter timescale $>10^{7}$~yr, thereby sustaining the gradual buildup of prebiotic molecules.

Observations reveal an inventory of complex organic molecules in some molecular clouds in the Milky Way’s CMZ. For example, the cloud G+0.693-0.027, located at a projected distance of about $\sim 100$ pc from the Galactic Center, contains nitriles, which are organic molecules with --C$\equiv$N groups and are precursors to RNA nucleotides \citep{2022FrASS...9.6870R}. This cloud is cold ($30$--$140$~K of gas temperature) and dense, with no ongoing star formation \citep{2024A&A...690A.121C}, making it an ideal environment for forming prebiotic molecules . Central clouds in the Milky Way likely have chemistry similar to that found in comets \citep{2019ARA&A..57..113A}, and many organics detected in comets and meteorites may have formed in these interstellar environments. According to the RNA world hypothesis, molecules like nitriles could have been delivered to early Earth by comets and meteors, jump-starting prebiotic chemistry \citep{2000A&A...354L...6C}. The Galactic Center may thus have contributed to the cosmic inventory of life's building blocks, demonstrating how a protected chemical niche can foster molecular complexity before planet formation.

The possible link between galactic core chemistry and the origin of life is speculative but worthy of consideration. Organic molecules essential for life likely have their origins in interstellar chemistry, once formed, they could be transported by galactic winds or incorporated into protoplanetary disks during star formation, thus seeding young planetary systems with prebiotic material. If LRDs were common in the early universe, their efficient dust and molecule production could have similarly enriched their surroundings. Subsequent galaxy mergers or hierarchical growth would then distribute these organics more broadly, potentially accelerating the onset of planet formation and prebiotic chemistry. The galaxies with low-luminosity, quiescent cores could act as reservoirs for complex chemistry, forming ``life-friendly'' environments within the otherwise hostile early universe.  LRDs could have contributed significantly to the cosmic budget of prebiotic molecules, providing the raw ingredients necessary for the later emergence of life.

\section{Outlook}
Direct observational verification of complex or prebiotic molecules in LRDs is challenging because of their cosmological distances, small physical sizes and masses, and low metallicities. However, these same conditions highlight the physical significance of LRDs: they plausibly represent the earliest proto-galactic nuclei to form, coeval with the first generations of stars, when heavy elements were scarce and only later increased through repeated cycles of star birth and death. They possibly host the formation of the first prebiotic molecules.

Moreover, recent JWST surveys indicate that LRDs may constitute a few$\times$10\% of galaxies in the early Universe \citep{2025arXiv251110725H}. Even if the molecular and dust production of any individual LRD is modest compared to that of the Milky Way today, their abundance and wide spatial distribution suggest that, in aggregate, they could have played a significant role in the early chemical enrichment. Through widespread dispersal of molecular precursors, these systems may have seeded subsequent generations of galaxies with the building blocks of complex and even prebiotic chemistry, potentially influencing the emergence of the chemical environments necessary for life in later cosmic epochs.

\section*{Acknowledgments}

We thank the Branch in the United States for supporting the computational resources used in this work. We are also grateful to the referee for the professional comments on astrochemistry, which helped improve the scientific discussion of the paper.

\bibliography{islands.bib}
\bibliographystyle{aasjournal} 

\end{document}